\documentclass[11pt]{article}
\usepackage{amssymb}
\usepackage{mathtools,extarrows}
\usepackage{amsmath,bm,amsfonts,color,amsthm, amssymb}
\usepackage[margin=2.5cm]{geometry}
\usepackage{graphicx}
\usepackage{rotfloat}
\usepackage[sort&compress]{natbib}
\usepackage{comment}
\bibpunct{(}{)}{;}{a}{}{,}
\usepackage[colorlinks,citecolor=blue,urlcolor=black]{hyperref}
\usepackage{lineno}
\usepackage{eurosym}
\usepackage{setspace}
\newcommand\numberthis{\addtocounter{equation}{1}\tag{\theequation}}
\usepackage{bbm, dsfont}

\usepackage{tikz}
\usetikzlibrary{decorations.pathmorphing} 
\usetikzlibrary{fit}         
\usetikzlibrary{backgrounds} 
\usetikzlibrary{positioning}
\usetikzlibrary{shadows}
\usepgflibrary{shapes}
\tikzstyle{observed}=[circle, thick, minimum size=0.7cm, draw=black!100, fill=black!20]
\tikzstyle{latent}=[circle, thick, minimum size=0.7cm, draw=black!100]
\tikzstyle{plate}=[rectangle, thick, inner sep=0.3cm, draw=black!100]
\tikzstyle{shadeplate}=[rectangle, thick, inner sep=0.3cm, draw=black!100, fill=black!10]

\numberwithin{equation}{section}
\theoremstyle{plain}

\usepackage{setspace}

\title{A Recommendation for Net Undercount Estimation in Iran Population and Dwelling Censuses}
\author{Sepideh Mosaferi \textsuperscript{a} \and
	Hamidreza Navvabpour\textsuperscript{b} }
	\date{
	\textsuperscript{a} Allameh Tabataba'i University,
Tehran, Iran \\
	\textsuperscript{b} Allameh Tabataba'i University and Statistical Research and Training Center, Tehran, Iran \\
	\ December 15, 2011}

\AtBeginDocument{\hypersetup{pdfborder={0 0 1}}}
	
\begin{document}

\maketitle

\begin{abstract}
Census counts are subject to different types of nonsampling errors. One of these main errors is coverage error. Undercount and overcount are two types of coverage error. Undercount usually occurs more than the
other, thus net undercount estimation is important. There are various methods for estimating the coverage error in censuses. One of these methods is dual system (DS) that usually uses data from the census and a post-enumeration survey (PES). In this paper, the coverage error and necessity of its evaluation, PES design and DS method are explained. Then PES associated approaches and their effects on DS estimation are illustrated and these approaches are compared. Finally, we explain the Statistical Center of Iran method of estimating net undercount in Iran 2006 population and dwelling census and a suggestion will be given for improving net undercount estimation in population and dwelling censuses of Iran.

\vspace{1cm}

\end{abstract}

\section{Introduction}
The fundamental purposes of the censuses are providing information to policy makers, planners, constructing statistical frames, etc. Thus an accurate and complete count of the population is important in censuses. Despite great efforts during the censuses, occurrence of errors such as coverage error is inevitable. Coverage error affects policy-making and planning issues of a country (see \cite{Freedman}). In addition, it leads to incomplete statistical frames, poor survey estimators, etc. To evaluate census performance many countries estimate coverage error after the censuses. Dual system (DS) is one of the methods that has been designed for this issue. Usually, information of census and post-enumeration survey (PES) are used in
the DS. Its fundamental was set by the United States Census Bureau (USCB) in the 1950s.

\cite{Hogan} discussed the accuracy of population census
counts by using post-enumeration survey and DS method in the United States (US). \cite{Hoganb} reviewed PES and DS. \cite{Kerr} compared procedures of estimating the net undercount in Canada, United States, Britain and Australia. \cite{USCBb} illustrated design and methodology of PES and DS in the US 2000 census. \cite{Dolson} explained the method of estimating coverage error in Canada. Iran estimated undercount in the 1976 census for the first time. Unfortunately there is not much information about the method has been used. \cite{Rohani} gave a proposal to the Statistical Center of Iran (SCI) to estimate net undercount in Iran 2006 Census. SCI estimated net undercount in Iran 2006 Census by using DS method. This method was slightly different from \cite{Rohani} recommendation.

On the other hand, net undercount is estimated with error and its results usually are not used to adjust the census count (\cite{Freedmanb}). Defects in census counts are assessed in order to conduct better censuses in the future.

In the second section of this paper, the method of estimating coverage
error and net undercount are explained and in section 3, PES is introduced. In the fourth section, DS estimation model is mentioned. Then, PES associated approaches with their effects on DS estimation are introduced and these approaches will be compared in the fifth section. The sixth section contains the method of estimating net undercount in Iran 2006 census. Finally, suggestion will be given for improving net undercount estimation in future population and dwelling censuses of Iran.

\section{Coverage Error}

Problems during a census may lead to errors in census counts. One of these errors is called coverage error. Two types of coverage error that is kind of nonsampling error, are undercount and overcount. In population censuses, some people may not be counted because of various reasons such as defects in census maps, errors in the field operations, etc. This lack of coverage is called undercount. Furthermore, some people may be counted more than once or some people that aren't in scope of the census may be counted. In this situation, overcount is occurred. In fact, gross coverage error $(G)$ is a combination of undercount $(U)$ and overcount $(O)$, i.e. $G=U+O$.

Net coverage error $(N)$ computes by subtracting overcount $(O)$ from undercount $(U)$, i.e. $N=U-O$. A positive number of this subtraction
indicates net undercount, while a negative number indicates net overcount. Usually undercount occurs more, and this subtraction will be positive. Let $T$ and $C$ be true population and counted population in a census, respectively. Then $T$ will be equal to $C+N$ (\cite{Dolson}).

Because of the importance of estimating net coverage error, many countries estimate true population according to one of the existing methods after censuses. Then, Net coverage error will be estimated by subtracting census count from true population estimation, i.e. $\hat{N}=\hat{T}-C$. Since, usually this subtraction is positive and represents net undercount, conventionally countries use the symbol $\hat{U}$ instead of $\hat{N}$, i.e. $\hat{U}=\hat{T}-C$. Also percent net
undercount estimation $(\hat{R})$ can be computed by $\hat{R}=\frac{\hat{U}}{\hat{T}} \times 100$.

\section{Post-enumeration Survey}
Most of the countries to evaluate accuracy of a census conduct a survey called post-enumeration survey (PES). In this survey, a representative sample of the census population is enumerated. Homeless persons and institutional households -such as nursing homes, prisons, and dormitories- usually are not counted in the PES, because assessing their errors is difficult and selection process is usually based on housing units/households. Thus, according to the country situation, a sample will be selected (see \cite{UN}). Then all households in the selected units/elements will be enumerated. Most of the countries use interviewer enumeration method instead of self-administrated
method to obtain the PES data. 

Sample from PES called population sample (P-sample). Information from P-sample can be used for assessing omissions that cause undercount in the census. Additionally, a sample of census enumerations in the same units/elements that the P-sample was selected is needed. This sample is called enumeration sample (E-sample). Information from E-sample uses for assessing erroneous enumerations in the census. Erroneous enumerations (e.g., duplicates, deaths before the census time, etc.)
cause overcount in the census. By matching data obtained from the two
sources of information, census omissions and erroneous enumerations can be determined. Overlapping of P-sample and E-sample reduces variance of DS estimate, fieldwork and cost (see \cite{USCBb}).

\section{Dual System Estimation}
DS method is a capture-recapture type technique. \cite{Petersen} used this method to estimate the population size of fish. The DS uses two sources of information. In our work, census is capture and PES is recapture. The methodology assumes a closed population, i.e. population size does not change during the study. Net undercount can be estimated by subtracting the true population estimation from DS method and the census count for different demographic groups such as sex, age groups and so forth that are called post-strata. Some countries use logistic regression rather than the post-stratification such as US in the 2010 census (see \cite{Moldoff}).
Most of developing countries use post-stratification. Therefore, we focus on post-stratification methodology in the present paper.

The DS estimation model can be applied within each post-stratum according to three following assumptions:
\begin{enumerate}
\item[a)] \textit{Homogeneity:} inclusion probability does not vary from one person to another in the census or PES.
\item[b)] \textit{Independence:} the probability of being one person in the PES does not depend on whether she/he was in the census. 
\item[c)] \textit{Perfect matching:} information of persons in two data sources (census and PES) must be matched without error.
\end{enumerate}

\noindent By using the obtained data from the E-sample and the P-sample, persons can be classified into four cells according to the presence or absence in the census and the PES (see Table \ref{table1}).

\begin{table}[ht]
\centering
\caption{Dual system estimation table} \label{table1}
\begin{tabular}{cccc}  
\hline\hline
 & \multicolumn{3}{c}{PES} \\
\cline{2-4}
Census & In & Out & Total \\
\hline
In & $X_{11}$ & $X_{10}$ & $X_{1+}$ \\
Out & $X_{01}$ & $X_{00}$ & $X_{0+}$ \\
Total & $X_{+1}$ & $X_{+0}$ & $T$ \\
\hline
\end{tabular}
\end{table}

All cell counts in Table 1 can be observed conceptually except for $X_{00}$. Persons in this cell do not enumerate in any surveys. If $\pi_{ij}$ is the inclusion probability for cell $(i, j)^{th}$ as $i, j = 0, 1$, according to the independence assumption $\pi_{ij}$ will be equal to $\pi_{i+} \pi_{+j}$ where $\pi_{i+}$ and $\pi_{+j}$ are marginal probabilities. Under the assumptions (a)-(c), a multinomial likelihood function can be considered, such that:
\begin{align*}
\label{eq:1}
L(T, \pi_{1+}, \pi_{+1})&=\frac{T!}{(T-X_{(1)})! X_{11}! X_{10}!} [\pi_{1+} \pi_{+1}]^{X_{11}} [\pi_{1+} (1-\pi_{+1})]^{X_{10}} \\
& \quad \times [(1-\pi_{1+}) \pi_{+1}]^{X_{01}} [(1-\pi_{1+}) (1-\pi_{+1})]^{T-X_{(1)}} \numberthis
\end{align*}

This likelihood function involves $T$, $\pi_{1+}$ and $\pi_{+1}$ as unknown parameters. Thus, we can write $L(T,\pi_{1+},\pi_{+1})$ in (\ref{eq:1}) as a product of two likelihood functions:
\begin{equation*}
L(T, \pi_{1+}, \pi_{+1}) = L_1(\pi_{1+},\pi_{+1}) \times L_2 (T,p^*),
\end{equation*}
where $p^∗=(1 - \pi_{1+})(1 - \pi_{+1})$ and
\begin{align*}
L_1(\pi_{1+}, \pi_{+1}) &= \frac{X_{(1)}!}{X_{11}! X_{10}! X_{01}!} 
 \times \frac{[\pi_{1+} \pi_{+1}]^{X_{11}} [\pi_{1+} (1-\pi_{+1})]^{X_{10}} [(1-\pi_{1+}) \pi_{+1}]^{X_{01}}}{[1-(1-\pi_{1+})(1-\pi_{+1})]^{X_{(1)}}},
\end{align*}
\begin{align*}
L_2(T,p^*)=\frac{T!}{X_{(1)}! (T-X_{(1)})!} [1-p^*]^{X_{(1)}}  [p^*]^{T-X_{(1)}}.
\end{align*}

$L_1(\pi_{1+},\pi_{+1})$ is a multinomial likelihood function, given the conditional distribution for the observable cells. $L_2(T,p^∗)$ is a binomial likelihood function. Maximum likelihood estimators of $\pi_{1+}$ and $\pi_{+1}$ can be obtained from $L_1(\pi_{1+}, \pi_{+1})$ and maximum likelihood estimator of $T$ can be obtained from $L_2(T,p^∗)$. These maximum likelihood estimators are (see \cite{Rao}, p. 38):
\begin{equation*}
\tilde{\pi}_{1+}=\frac{X_{11}}{X_{+1}}, \quad \tilde{\pi}_{+1}=\frac{X_{11}}{X_{1+}}, \quad \tilde{T}=\frac{X_{1+} X_{+1}}{X_{11}}.
\end{equation*}

$\tilde{T}$ is called DS estimator. The quantities of $X_{+1}$ and $X_{11}$ in estimator $\tilde{T}$ are unknown because information from PES is available from P-sample not from whole the population. Also $X_{1+}$ is unknown and differs from actual census counts due to erroneous enumerations or other errors (\cite{Mulry}). In practice, true population can be estimated by the following equation that is called empirical DS estimator, too:
\begin{equation}
\label{eq:2}
\hat{T}=\frac{\hat{X}_{1+} \hat{X}_{+1}}{\hat{X}_{11}}=(C - II) (1-\frac{\hat{EE}}{\hat{N}_e}) \frac{\hat{N}_p}{\hat{M}}
\end{equation}
In (\ref{eq:2}), $\hat{X}_{1+}$ is $(C - II) (1-\frac{\hat{EE}}{\hat{N}_e}),$ where
\begin{enumerate}
\item[] $C:$ census count,
\item[] $II:$ number of whole-person census imputations,
\item[] $\hat{EE}:$ weighted estimate of E-sample erroneous enumerations, and
\item[] $\hat{N}_e:$ weighted E-sample total.
\end{enumerate}
\noindent Also in (\ref{eq:2}) $\hat{X}_{+1}$ and $\hat{X}_{11}$ are $\hat{N}_p$ and $\hat{M}$, respectively, where
\begin{enumerate}
\item[] $\hat{N}_p:$ weighted P-sample total, and
\item[] $\hat{M}:$ weighted estimate of P-sample matches.
\end{enumerate}

To clarify more, $\hat{X̂}_{1+}$ points out an estimate of population who are correctly enumerated in the census (where erroneous enumerations and whole person census imputations must be factored out). Records of whole-person imputation are not included in the matching process because the executive office of the census imputes almost their whole characteristics (\cite{Nash} and \cite{Dolson}). $\hat{X}_{+1}$ is an estimate of the total number of people counted in the PES and $\hat{X}_{11}$ is an estimate of the total number of people counted in
both of the census and the PES (see \cite{Hoganc}). In equation \ref{eq:2}, $\frac{\hat{X}_{+1}}{\hat{X}_{11}}$ (or $\frac{\hat{N}_p}{\hat{M}}$) is called inverse of match rate, because $\frac{\hat{X}_{11}}{\hat{X}_{+1}}$ is referred to the proportion estimation of P-sample persons who can be matched to the ones enumerated in the census.

To estimate the quantities in (\ref{eq:2}), countries apply alternative sample designs and suitable weights according to the design. Some countries use the second part of equation (\ref{eq:2}) $[(C-II)(1-\frac{\hat{EE}}{\hat{N}_e})\frac{\hat{N}_p}{\hat{M}}]$ slightly differ. For example, some countries do not do any whole-person census imputations or some other countries just use the information of the E-sample for estimating the weighted total number of people counted correctly in the census ($\hat{X}_{1+}$) instead of estimating a correct enumeration factor from the E-sample and then multiply it by the census count.

Under the assumptions (a)-(c), the odds ratio in Table \ref{table1} is equal to 1, thus $\hat{X}_{00}=\frac{\hat{X}_{10} \hat{X}_{01}}{\hat{X}_{11}}$. On the other hand by using (\ref{eq:2}), $\hat{T}$ can be rewritten as follows:
\begin{align*}
\label{eq:3}
\hat{T}&=\frac{\hat{X}_{1+} \hat{X}_{+1}}{\hat{X}_{11}}=\frac{(\hat{X}_{11}+\hat{X}_{01})}{\hat{X}_{11}} \\
& = \hat{X}_{11}+\hat{X}_{10}+\hat{X}_{01}+\frac{\hat{X}_{10} \hat{X}_{01}}{\hat{X}_{11}} = \hat{X}_{11}+\hat{X}_{10}+\hat{X}_{01}+\hat{X}_{00} \\
& = \hat{X}_{(1)}+\hat{X}_{00}. \numberthis
\end{align*}
Some countries use equation (\ref{eq:3}) for true population estimation.

\section{PES Approaches and Their Effects on DS Estimation}
Countries try to conduct PES immediately after finishing the enumeration
and field operations of censuses, but in this period of time, some people may move from their places at the time of conducting the census. We classify persons to four categories at the time of PES as follows:
\begin{enumerate}
\item[i)] \textit{Non-mover:} A person who resided in the same household at the time of PES and census.
\item[ii)] \textit{In-mover:} A person who resided in the household at the time of PES but did not reside in the household at the time of census.
\item[iii)] \textit{Out-mover:} A person who resided in the household at the time of census but did not reside in the household at the time of PES.
\item[iv)] \textit{Out-of-scope:} A person who does not belong to the target population of census, such as a child born after the census time.
\end{enumerate}

We consider a closed population during this study. Out-of-scopes do not
consider in the study. Movers (in-movers or out-movers) must be suitably
processed and used in the estimators of $N_p$ and $M$. If their records are not assessed, part of the target population of the census has been spuriously excluded from the PES and coverage error estimation would be biased. In addition, it is important to consider a person is classified as matched one during the matching operations between the census and PES when she/he is counted with similar information in both sources, whether moved or not moved (\cite{UN}, chapter 7).

There are three basic procedures for evaluating census coverage in a PES. The procedures differ in the treatment of movers (\cite{USCB}). These procedures are known as A, B and C that can influence PES questionnaire, matching operations and empirical DS estimate.

\subsection{Procedure A}
During the PES data collection, all persons are identified that lived in the sample households at the census time. These persons are non-movers and out-movers. Out-movers in a household can be reported by other members of the household but when all members of the household have moved out, they may be reported by proxy respondents (such as neighbors). This collected information in the PES must be matched to the census information in the sample areas. Then separate estimates of the number of non-movers, out-movers and matched non-movers and matched out-movers can be made.

By using this method, $\frac{\hat{N}_p}{\hat{M}}$ in the empirical DS estimator (\ref{eq:2}) is employed
\begin{equation*}
\frac{\hat{N}_p}{\hat{M}}=\frac{\hat{N}_{\text{non}}+\hat{N}_{\text{out}}}{\hat{M}_{\text{non}}+\hat{M}_{\text{out}}},
\end{equation*}
where $\hat{N}_{\text{non}}$ is the estimate of non-movers from P-sample, $\hat{N}_{\text{out}}$ is the estimate
of out-movers from P-sample, $\hat{M}_{\text{non}}$ is the estimate of non-mover matches from P-sample and $\hat{M}_{\text{out}}$ is the estimate of out-mover matches from P-sample.

\subsection{Procedure B}
During the PES data collection, all current persons are identified that live in the sample households at the PES time. These persons are non-movers and in-movers. Thus in-movers are enumerated at their new addresses in the PES. Since people respond for themselves; hence, the field data collection operation is easier and more complete than procedure A, but the current places of living in-movers are different from their census places. The census addresses for in-movers must be reported by respondents. Thus, matching operation for in-movers is difficult and involves searching in the areas where they were during
the census that may be far from the sample areas in the PES. Then separate estimates of the number of non-movers, in-movers, matched non-movers and matched in-movers can be made.

By using this method, $\frac{\hat{N}_p}{\hat{M}}$ in the empirical DS estimator (\ref{eq:2}) is employed
\begin{equation*}
\frac{\hat{N}_p}{\hat{M}}=\frac{\hat{N}_{\text{non}}+\hat{N}_{\text{in}}}{\hat{M}_{\text{non}}+\hat{M}_{\text{in}}},
\end{equation*}
where $\hat{N}_{\text{in}}$ is the estimate of in-movers from P-sample and $\hat{M}_{\text{in}}$ is the estimate of in-mover matches from P-sample.

\subsection{Procedure C}
During the PES data collection, all current persons are identified in the sample households (non-movers and in-movers) and persons that lived in the sample households just in the census (out-movers). This method is a combination of procedures A and B, because estimates of the number of
non-movers and movers (by using the in-movers) come from procedure B
and estimates of the matched non-movers and matched movers (by using the
out-movers) come from procedure A. Thus, separate estimates of the number of non-movers, in-movers, out-movers, matched non-movers and matched out-movers must be made in this method.

By using this method, $\frac{\hat{N}_{p}}{\hat{M}}$ in the empirical DS estimator (\ref{eq:2}) is employed as follows:
\begin{equation*}
\frac{\hat{N}_p}{\hat{M}}=\frac{\hat{N}_{\text{non}}+\hat{N}_{\text{in}}}{\hat{M}_{\text{non}}+\frac{\hat{M}_{\text{out}}}{\hat{N}_{\text{out}}}\hat{N}_{\text{in}}}.
\end{equation*}
In procedure C, the number of matched in-movers cannot be estimated directly. According to the closed population assumption, in-movers and out-movers are in a same group which is called movers. The total number of in-movers is equivalent to the total number of out-movers at the national level (however, it may differ in post-stratum levels because of alternative reasons). Thus, the match rate of in-movers is equivalent to the match rate of out-movers ($\frac{\hat{M}_{\text{out}}}{\hat{N}_{\text{out}}}$). Hence, the total number of matched in-movers can be estimated indirectly by ($\frac{\hat{M}_{\text{out}}}{\hat{N}_{\text{out}}} \hat{N}_{\text{in}}$).

\subsection{Comparisons among Procedures A, B and C}
A main difference between procedures A and B involves movers. In procedure A, movers are estimated by out-movers. If proxy respondents do not report out-movers suitably, the number of movers will be underestimated. Thus, this method may be leading to underestimation of the census omissions. Matching operations in procedure A are easy and cheap, because the search is limited to the sample areas. In procedure B, movers are estimated by in-movers. The field operations in procedure B are easier than procedure A, because current persons in the sample areas are asked by themselves. Thus, a better estimate of movers can be provided in procedure B. Matching operations for in-movers in procedure B are difficult, specifically when the respondents do not give suitable information about the addresses of in-movers
at the census time, which sometimes lead to overestimation of the census
omissions. Procedure B is expensive because the matching operations are
not limited to the sample areas. Hence this method is difficult for countries without enough equipment that may be leading to heavy bias in matching. Procedure C uses the advantages of matching operations of procedure A and field operations of procedure B by using in-movers for estimating the movers. Procedure C is more expensive than procedure A, because all persons at the time of PES and census must be identified. Procedure C is cheaper than procedure B, because procedure C does not follow B matching operations.

Procedure A is the weakest method, because only persons who lived at the
census time in sample areas will be assessed during the PES. Assessment of the non-movers and out-movers by this method can be increased the number of doubly-missing persons because persons who have been missed in the census, tend to be missed in PES, too. Therefore, this method leads to underestimation of true population. Procedure B provides a better estimate of movers but for the difficulties in matching, it can be used in countries with enough equipment (e.g. enough budgeting, computer-assisted matching, adequately addressing system in urban and rural areas and so forth). US used this procedure in 2010 PES (\cite{Moldoff}). Procedure C has advantages of other procedures.

\section{Method of Net Undercount Estimation in Iran 2006 Census}
Post-enumeration survey was conducted ten days after the Iran 2006 census and lasted five days. The main purpose of conducting PES was estimating the percent of net undercount at national and subnational (urban, rural areas and provinces) levels. The target population of this survey was all usual residential households in urban and rural areas except moving households (homeless persons and nomads) and institutional households.

The PES was based on two stage cluster sampling design in which primary sampling unit (PSU) was each district, and secondary sampling unit
(SSU) was each usual residential household. The PES sampling frame was a
list of all districts in geographical file that was used for the 2006 census. Thus at the first stage, sufficient number of sample districts was selected systematically in urban and rural areas of each province (for detailed information on the sample size and allocation to the provinces, see \cite{SCIa}). At the second stage, 50 usual residential households of selected districts in urban areas and 100 usual residential households of selected districts in rural areas were
selected. SCI couldn't select households directly within the sampled PSUs because of the frame imperfection of households in PES sampling frame and the burden of field operation. For these reasons, SCI selected just the start point of field operation at random. During the PES field operation under the assumption of random distribution of households in consecutive housing units and for the easiness of matching operations, enumerators listed households from southwest of the start point in anticlockwise direction (opposite direction employed in the census) until the specified number of residential households was listed. These households might be present or temporarily absent\footnote{1: see Remarks} at the time of PES.

In order to have the perfect independence between the PES and the
census and better assessing errors, households were listed again during the PES. Also PES enumerators were not employed as census enumerators in
the same areas. Each household was listed in a special listing form. In this form, some information of households such as characteristics of their place of living (complete address, place number and so forth) and head of households were gathered. In addition, enumerator assessed “Does the household move into the place after the census time?”. When there was a positive response, enumerator gathered information of household who had resided in the place at the census time from proxies. This information was used for the matching operation of households between the PES and census (Note that there were similar forms for listing households in the census). Then a face-to-face interview (the same as the census) was done for the present listed households by enumerators to complete PES questionnaire. Interview of households was
done simultaneously with the listing operation by the same enumerators.

The PES questionnaire consisted of some demographic questions and
questions aimed at identifying movers (both in-movers and out-movers) and out-of-scopes (such as a child born after the census time). Also persons died after the census time was assessed because they were included in the census target population (also could be treated as out-movers). Thus during the PES, all non-movers, in-movers, out-movers and out-of-scopes were assessed. This information was used for matching operation of persons between the PES and census.

After the data collection, manual matching operations were done by
clerks, supervisors and professional staffs to determine match status. Matching was a comparison between records of households and persons in the PES with records from the census in the same sample areas. Matching operations were carried out in two phases. During the initial matching phase, households and persons were matched. Those households and persons that their matching situations could not be specified were followed-up by telephone or fieldwork. Then final matching phase was done.

The household matching involved determination whether every house-
hold in the PES had been counted in the census. It involved searching
through census information to locate the addresses of the PES households. After household matching, person matching was conducted within matched households. The information of persons such as name, age, sex, etc. was compared between the census and the PES to determine the match status.
Finally, a field match code was recorded for each status (\cite{SCIb}). Different statuses of household and person in the initial matching are shown in Tables \ref{table2} and \ref{table3}, respectively. Then all households and persons with codes 40 and 50 in the initial matching were sent to follow-up. When data from following-up were available, final matching was done. Results are shown in Table \ref{table4}.

According to the matching operations and information from Tables \ref{table2}-\ref{table4},
true population was estimated by DS method using equation (\ref{eq:3}) as follows:
\begin{align*}
\label{eq:4}
\hat{T} &=\hat{f}_{10}+\hat{f}_{30}+(\hat{f}_{42/1}+\hat{f}_{42/2}+\hat{f}_{42/3}+\hat{f}_{42/4}) \\
& \quad + (\hat{f}_{52/1}+\hat{f}_{52/2}+\hat{f}_{52/3}+\hat{f}_{52/4})+\hat{N}_{22}, \numberthis
\end{align*}
where $\hat{X}_{(1)}=\hat{f}_{10}+\hat{f}_{30}+(\hat{f}_{42/1}+\hat{f}_{42/2}+\hat{f}_{42/3}+\hat{f}_{42/4})+ (\hat{f}_{52/1}+\hat{f}_{52/2}+\hat{f}_{52/3}+\hat{f}_{52/4})$ and $\hat{X}_{00}=\hat{N}_{22}$. Symbols in equation (\ref{eq:4}) were defined as: 
\begin{enumerate}
\item[] $\hat{f}_{10}:$ weighted estimate of matched persons,
\item[] $\hat{f}_{30}:$ weighted estimate of out-mover/dead persons,
\item[] $\hat{f}_{42/1}:$ weighted estimate of omissions in the census because the place of household was omitted in the census,
\item[] $\hat{f}_{42/2}:$ weighted estimate of omissions in the census because the household was not recognized in the place in the census,
\item[] $\hat{f}_{42/3}:$ weighted estimate of omissions in the census because the number of households was falsely recognized in the place in the census,
\item[] $\hat{f}_{42/4}:$ weighted estimate of omissions in the census in matched household and persons in household with code 42,
\item[] $\hat{f}_{52/1}:$ weighted estimate of omissions in the PES because the place of household was omitted in the PES,
\item[] $\hat{f}_{52/2}:$ weighted estimate of omissions in the PES because the household was not recognized in the place in the PES,
\item[] $\hat{f}_{52/3}:$ weighted estimate of omissions in the PES because the number of households was falsely recognized in the place in the PES,
\item[] $\hat{f}_{52/4}:$ weighted estimate of omissions in the PES in matched households, and
\item[] $\hat{N}_{22}:$ weighted estimate of omissions in both of census and PES.
\end{enumerate}
Also $N_{22}$ was estimated using (\ref{eq:5}) as follows (see \cite{SCIa}):
\begin{equation}
\label{eq:5}
\hat{N}_{22}=\frac{(\hat{f}_{42/1}+\hat{f}_{42/2}+\hat{f}_{42/3}+\hat{f}_{42/4}) \times (\hat{f}_{52/1}+\hat{f}_{52/2}+\hat{f}_{52/3}+\hat{f}_{52/4})}{\hat{f}_{10}}
\end{equation}

In above statements, weighted estimate of specific status at the specified group level (such as a province) was computed by $\hat{f}_{S}=\sum_{s \in S}w_s$, where $S=$ set of persons with specific status at specified group level, $s=$ each person in set $S$ and $w_s=$ weight of $s$th person. This weight was computed according to the
sample design. Whereas two stage cluster sampling design was used and all persons in households were enumerated; the weight of persons was equal to the weight of households and was inverse of the selection probability for one household in rural or urban areas of the provinces.
This selection probability was computed by $\frac{n.d}{tn.d} \times \frac{n.h}{tn.h}$ where $n.d=$ number of sampled districts in urban or rural areas of a province, $tn.d=$ total number of districts in urban or rural areas of a province, $n.h=$ number of sampled households in specific district, and $tn.h=$ total number of households in the specific district. Finally, percent net undercount estimation ($\hat{R}$) was computed by $\hat{R}=\frac{\hat{U}}{\hat{T}} \times 100$ at specified group levels. To compute $\hat{U}$ by $\hat{T}-C$ corresponding census counts of moving households and institutional households were extracted from initial census
counts, since the PES target population just covered the usual residential household population.

\subsection{Analysis of the Method of Net Undercount Estimation
in Iran 2006 Census}
In this subsection, we analyze the method of net undercount estimation and try to match this method to A, B, or C methods.
United Nations (UN) recommends countries to consider one of the A, B
or C methods to conduct PES and use PES and census information in DS
model to estimate the net undercount (see, \cite{UN}). This recommendation was not adopted adequately by SCI for estimating the net undercount in 2006 census. 

If we want to classify Iran 2006 PES in one of the A, B, or C categories, we will encounter some difficulties. During the PES, demographic information of all persons (non-movers, out-movers and in-movers) were asked. SCI could not match in-movers, because the search areas were limited to the sample areas. Therefore, we could not classify PES in category B. Since in-movers (persons/households with code 20 of course without births) were not used in $\hat{T}$ for estimating in-movers, the PES can not be classified in category C. SCI
matched non-movers and out-movers because of the usage of sample areas as search areas. 

\begin{sidewaystable}[H]
\centering
\caption{Household Matching Results in the Initial Matching} \label{table2}
\begin{tabular}{llll}  
\hline\hline
 & \multicolumn{3}{c}{PES} \\
\cline{2-4}
& \multicolumn{2}{l}{Household was listed} & Household was \\
\cline{2-3}
Census & Temporarily absent & With questionnaire & not listed \\
\hline
Household was listed & $\#$ Household & Code 42 was assigned for household & $\S$Household was removed \\
Without questionnaire & from matching questionnaire & and code 42/4 for its members & from matching operation \\
& and estimate of $T$ & & and estimate of $T$ \\
\hline
Household was listed & Code 10 was assigned & Code 10 was assigned for household: & -Thorough acquired information from \\
With questionnaire & for both of households & matched household & proxies, household was recognized \\
& and its members & & as out-mover with the similar characteristics \\
& & & in census and PES: Code 30 \\
& & & -According to the available information, \\
& & & household did not move out of the sample \\
& & & area after census time: Code 50 \\
\hline
Household was not listed & $\P$ Household was removed & -According to the avaiable information, & \\
& from matching operation & household moved into the sample area & \\
& and estimate of $T$ & after census time: Code 20 & \\
& & -According to the available information, & \\
& & household did not move into the sample & \\
& & area after census time: Code 40 & \\
\hline
\end{tabular}
Note: Households with code 20 did not use to estimate $T$. 
\end{sidewaystable}

\begin{table}[ht]
\centering
\caption{Person matching results in the initial matching} \label{table3}
\begin{tabular}{ll}  
\hline\hline
Description & Outcome code\\
\hline
Person who was non-mover and enumerated in & Matched person: Code 10 \\
census and PES with similar characteristics that & \\
could be referred to the same one & \\
& \\
Person who was enumerated in PES but not in & In-mover/Born person: Code 20 \\
census and born or moved into the sample area & \\
after census time & \\
& \\
Person who was enumerated in PES but not in & Code 40 \\
census and was not reported as birth or in-mover & \\
after census time & \\
& \\
Person who was enumerated in census and was & Out-mover/Dead person: Code 30 \\
reported as out-mover or death in PES with & \\
the similar characteristics in both sources & \\
& \\
Person who was enumerated in census but not & Code 50 \\
in PES & \\
\hline
\end{tabular}

Note: Persons with code 20 did not use to estimate $T$.
\end{table}

In addition, different statuses of non-movers and out-movers were considered and used in the estimate of $T$; however, we can hardly classify the 2006 PES in category A, because of the existence of following problems:
\begin{itemize}
\item According to available information from SCI and Tables \ref{table2} and \ref{table3}, persons were defined as matched when they lived in a same place in PES and in the census and their demographic information were similar in both sources (\cite{SCIb}). It is noteworthy to consider that people with code 30 referred to persons who lived and counted in the sample area during the census and were reported as out-movers or deaths by other members of household or by proxies during the PES, and information in both sources were similar. Conducting this operation by SCI is a kind of matching, because when information from both sources was similar, code 30 was assigned. These people must be identified as matched out-movers. SCI definition of matched persons was limited to non-movers. Therefore, unsuitably assessing the out-movers, improperly defining their statuses and omitting them from part of the true population led to bias in undercount estimation.
\item Improper usage of $\hat{f}_{30}$ in $\hat{T}$ led to bias. This bias points out that $\hat{f}_{30}$ ($\hat{f}_{30} \geq 0$, because of referring to the number of enumerations) was omitted from (\ref{eq:5}) to estimate $N_{22}$. Thus, it is not clear that persons
with code 30 should be considered in which cells of Table \ref{table1}. Note to the following possible situations:
\begin{enumerate}
\item If we assume that persons with code 30 were in cell $X_{10}$, we can conclude that:
\begin{itemize}
\item[] $\hat{X}_{11}=\hat{f}_{10},$
\item[] $\hat{X}_{10}=\hat{f}_{30}+\hat{f}_{52/1}+\hat{f}_{52/2}+\hat{f}_{52/3}+\hat{f}_{52/4}$ and
\item[] $\hat{X}_{01}=\hat{f}_{42/1}+\hat{f}_{42/2}+\hat{f}_{42/3}+\hat{f}_{42/4},$ so
\item[] $\hat{X}_{00}=\hat{N}_{22}= (\hat{f}_{42/1}+\hat{f}_{42/2}+\hat{f}_{42/3}+\hat{f}_{42/4}) \times (\hat{f}_{52/1}+\hat{f}_{52/2}+\hat{f}_{52/3}+\hat{f}_{52/4})/\hat{f}_{10}$
\end{itemize}
\item[] Therefore, omitting $\hat{f}_{30}$ from the numerator of $\hat{N}_{22}$ leads to underestimate the true population and net undercount.
\item Else if we assume that persons with code 30 were in cell $X_{11}$, we can conclude that:
\begin{itemize}
\item[] $\hat{X}_{11}= \hat{f}_{10}+ \hat{f}_{30},$
\item[] $\hat{X}_{10}= \hat{f}_{52/1} + \hat{f}_{52/2} + \hat{f}_{52/3} + \hat{f}_{52/4}$ and
\item[] $\hat{X}_{01}= \hat{f}_{42/1} + \hat{f}_{42/2} + \hat{f}_{42/3} + \hat{f}_{42/4}$, so
\item[] $\hat{X}_{00}=\hat{N}_{22}=\frac{(\hat{f}_{42/1} + \hat{f}_{42/2} + \hat{f}_{42/3} + \hat{f}_{42/4}) \times (\hat{f}_{52/1} + \hat{f}_{52/2} + \hat{f}_{52/3} + \hat{f}_{52/4})}{\hat{f}_{10}+\hat{f}_{30}}$.
\end{itemize}
Therefore, omitting $\hat{f}_{30}$ from the denominator of $\hat{N}_{22}$, leads to overestimation in true population and net undercount. As mentioned before, assigning code 30 by SCI is a kind of matching
process (matched out-movers). Therefore, using $\hat{f}_{30}$ in the denominator of $\hat{N}_{22}$ is more appropriate.
\end{enumerate}
\end{itemize}

According to Table \ref{table2}, in the cell marked $\#$ (households who were listed but temporarily absent in the PES and were listed but without questionnaires in the census), households were removed from matching operations and estimating $T$. These households were identified in both surveys but because of different reasons such as temporary absence or refusal to participate were not interviewed and there were no questionnaires for them. Omitting these households from non-movers leads to bias. Thus, for these households nonresponse, noninterview adjustment method must be applied. The noninterview adjustment method spread weight of these households among households that were interviewed in the same noninterview adjustment cell.

Noninterview adjustment cells could be made according to the households'
characteristics, such as types of basic address (single unit, multiunit such as apartments, etc.), in each sample area. In the cell marked $\S$ (households who were not listed in PES and were listed but without questionnaires in the census), households were removed from matching operations and estimating $T$. For these households, there were no questionnaires in the census may be because of the refusal of participation and they were not listed in PES because of the error in fieldwork by enumerator or moving out of the sample
area. Omitting these households leads to underestimate the non-movers or
out-movers.

It was better to consider these households for following-up and a
more suitable status should be considered for them by acquired information. Also In the cell marked $\P$ (households who were listed but temporarily absent in the PES and were not listed in the census), households were removed from matching operations and estimating $T$. For these households, there were no questionnaires in PES because of temporary absence, and they were not listed in the census because of an error in fieldwork or moving into the sample area. Omitting these households leads to underestimate the non-movers or
in-movers. It was better to consider these households for following-up and a more suitable status should be considered for them by acquired information. 

\begin{table}[ht]
\centering
\caption{Person and Household matching results in the final matching} \label{table4}
\begin{tabular}{ll}  
\hline\hline
Status of household or person in initial & Possible status of household or person \\
matching & after follow-up \\
\hline
Person who was enumerated in PES but not & - Overcount in PES: Code 41 \\
in census and was not reported as birth or in- & - Undercount in census: Code 42/4 \\
mover after census time (Code 40) & \\
& \\
Person who was enumerated in census but & - Overcount in census: Code 51 \\
not in PES (Code 50) & - Undercount in PES: Code 52/4 \\
& \\
Household who was listed in PES but was & - Overcount in PES: Code 41 \\
not listed in census and did not move into the & - Omission the place of household in \\
sample area after census time (Code 40) & census: Code 42/1 \\
& - No recognition of the household in \\
& place by enumerator in census: Code 42/2 \\
& - False recognition of the number of \\
& households by enumerator in census: \\
& Code 42/3 \\
& \\
Household who was listed in census but was & - Overcount in census: Code 51 \\
not listed in PES and did not move out of the & - Omission the place of household in \\
sample area after census time (Code 50) & PES: Code 52/1 \\
& - No recognition of the household in \\
& place by enumerator in PES: Code 52/2 \\
& - False recognition of the number of \\ 
& households by enumerator in PES: \\
& Code 52/3 \\
\hline
\end{tabular}

Note: Persons and households with codes 41 and 51 did not use to estimate $T$ because of erroneous enumerations in census or PES.
\end{table}

\section{A Recommendation for Net Undercount Estimation in Future Censuses}
During the planning for net undercount estimation by conducting PES and
using DS method in a country, it is appropriate to choose one of the standard methods A, B and C according to country situation. Designing a suitable questionnaire without extra questions to identify movers is crucial.

To recommend an appropriate method for the case of Iran, we must note
that procedure A is unsuitable because respondents may not report out-
movers adequately. Also assessing out-movers and non-movers leads to higher doubly-missing and underestimation of population. Procedure B provides a better estimate of movers by using in-movers than procedure A. In-mover matching is difficult and costly that exacerbates problems, which involve searching for in-movers in the places where they were enumerated in the census. In addition, there is not a reliable addressing system for some regions like rural areas in Iran. Therefore, when in-movers live in the regions with unreliable addressing system during the census where they are far from the sample areas in PES, matching clerks have to search all of census information
for those regions to determine their statuses. This activity takes a long time and yields a high level of unresolved cases. So procedure B is not a proper choice for Iran.

Procedure C is a combination of procedures A and B which takes the
advantages of both procedures to reduce matching difficulties and improve the estimation of movers. Estimating the number of movers by in-movers is more reliable since information is collected from in-movers themselves. Also estimating match rate of movers by out-movers is more accurate and avoids the difficulties of in-movers matching. For these reasons, we recommend procedure C to estimate net undercount for the case of Iran.

In accordance with procedure C, the non-movers, in-movers, out-movers,
deaths and births must be assessed in sample areas during the PES data
collection. Therefore, PES questionnaire should be consisted of suitable
questions to discern these persons. When all members of a household are
recognized as in-movers in a place during the data collection in PES, it is better to ask current household or neighbors about former household (plus its characteristics) lived in the place during the census time. This activity helps to identify out-movers accurately and reduces follow-up workloads. After data collection, information of out-of-scopes (such as births) is removed and is not used to estimate $T$.

Moreover, in-movers (persons who move into the sample areas after the census time) are specified from PES information to estimate in-movers. Then during the matching operations, census records will be matched with PES records to search out-movers (plus deaths) and non-movers in sample areas to classify them as matched ones or other suitable statuses. In this manner, persons who enumerated in one source and
missed in the other source will be identified. When matching operation and following-up were finished, true population will be estimated. Hence we must have the following estimates (weighted estimate from the sample):
\begin{enumerate}
\item[a:] Estimate of the total number of non-movers from P-sample.
\item[b:] Estimate of the total number of out-movers from P-sample.
\item[c:] Estimate of the total number of in-movers from P-sample.
\item[d:] Estimate of the total number of matched non-movers.
\item[e:] Estimate of the total number of matched out-movers.
\item[f:] Estimate of the total number of matched in-movers indirectly by $[(e/b)c]$.
\item[g:] Estimate of correctly enumerated population in census by using the information of census at the sample area levels and results of matching.
\end{enumerate}

\begin{table}[ht]
\centering
\caption{Dual system estimation table} \label{table5}
\begin{tabular}{ccc}  
\hline\hline
 & \multicolumn{2}{c}{PES} \\
\cline{2-3}
Census & In & Out \\
\hline
In & $\hat{X}_{11}=d+f$ & $\hat{X}_{10}=\hat{X}_{1+}-\hat{X}_{11}=g-(d+f)$ \\
Out & $\hat{X}_{01}=\hat{X}_{+1}-\hat{X}_{11}=(a+c)-(d+f)$ & $\hat{X}_{00}=\frac{\hat{X}_{10} \hat{X}_{01}}{\hat{X}_{11}}$ \\
\hline
\end{tabular}
\end{table}
Table \ref{table5} is provided according to above available estimates.
True population ($T$) can be estimated by one of the equations (\ref{eq:2}) or (\ref{eq:3}). The results of these equations are similar. We introduce $\hat{T}$ in (\ref{eq:2}) by $\hat{T}_1$ and
in (\ref{eq:3}) by $\hat{T}_2$. Now we show $\hat{T}_2=\hat{T}_1$ as follows:
\begin{align*}
\hat{T}_2 & =\hat{X}_{11}+\hat{X}_{10}+\hat{X}_{01}+\hat{X}_{00}=(d+f) + g - (d+f) \\
& \quad + (a+c)-(d+f)+\frac{[g-(d+f)] \times [(a+c)-(d+f)]}{d+f} \\
& = g+(a+c)-(d+f) + \frac{[g-(d+f)] \times [(a+c)-(d+f)]}{d+f} \\
& =\frac{[g+(a+c)] \times [d+f] - [d+f]^2 + [g-(d+f)] \times [(a+c)-(d+f)]}{d+f}\\
& = \frac{g(a+c)}{d+f}=\frac{\hat{X}_{1+} \hat{X}_{+1}}{\hat{X}_{11}}=\hat{T}_1.
\end{align*}

\section{Conclusion}
In this paper we considered the coverage error as one of the most important nonsampling errors in the census. As mentioned before undercount may be occurring more than overcount. PES along with DS method to estimate the net undercount were explained and three basic procedures (A, B and C) for PES were introduced. Finally according to the Iran situation, procedure C was recommended among other methods to estimate net undercount in Iran future censuses.

\newpage
\clearpage

\vspace{0.5cm}

\noindent \textbf{Acknowledgement}

\noindent The authors would like to thank the Statistical Center of Iran that provides information about the census and PES conducted in 2006.

\vspace{0.5cm}

\noindent \textbf{Remarks}

\noindent 1. Temporarily absent households are referred to households who are not at homes during the PES time for all reasons except migration.

\bibliographystyle{ims}
\bibliography{Bibliography}


\end{document}